\documentclass[aps,prl,reprint,amsmath,amssymb,superscriptaddress,nobibnotes]{revtex4-1}
\usepackage{graphicx,SIunits}
\usepackage{bm}     
\usepackage{hyperref} 
\usepackage{dsfont}    
\usepackage{color}

\begin{document}

\title{Experimental implementation of arbitrary entangled operations}

\author{Seongjin Hong}
\affiliation{Center for Quantum Information, Korea Institute of Science and Technology (KIST), Seoul, 02792, Korea}
\affiliation{Institute of Physics and Applied Physics, Yonsei University, Seoul 03722, Korea}

\author{Chang Hoon Park}
\affiliation{Center for Quantum Information, Korea Institute of Science and Technology (KIST), Seoul, 02792, Korea}
\affiliation{Department of Electrical and Computer Engineering, Ajou University, Suwon 16499, Korea}

\author{Yeon-Ho Choi}
\affiliation{Center for Quantum Information, Korea Institute of Science and Technology (KIST), Seoul, 02792, Korea}
\affiliation{Division of Nano and Information Technology, KIST School, Korea University of Science and Technology, Seoul 02792, Korea}

\author{Yong-Su Kim}
\affiliation{Center for Quantum Information, Korea Institute of Science and Technology (KIST), Seoul, 02792, Korea}
\affiliation{Division of Nano and Information Technology, KIST School, Korea University of Science and Technology, Seoul 02792, Korea}

\author{Young-Wook Cho}
\affiliation{Center for Quantum Information, Korea Institute of Science and Technology (KIST), Seoul, 02792, Korea}

\author{Kyunghwan Oh}
\affiliation{Institute of Physics and Applied Physics, Yonsei University, Seoul 03722, Korea}

\author{Hyang-Tag Lim}
\email{hyangtag.lim@kist.re.kr}
\affiliation{Center for Quantum Information, Korea Institute of Science and Technology (KIST), Seoul, 02792, Korea}

\date{\today} 

\begin{abstract}

Quantum entanglement lies at the heart of quantum mechanics in both fundamental and practical aspects. The entanglement of quantum states has been studied widely, however, the entanglement of operators has not been studied much in spite of its importance. Here, we propose a scheme to realize arbitrary entangled operations based on a coherent superposition of local operations. Then, we experimentally implement several intriguing two-qubit entangled operations in photonic systems. We also discuss the generalization of our scheme to extend the number of superposed operations and the number of qubits. Due to the simplicity of our scheme, we believe that it can reduce the complexity or required resources of the quantum circuits and provide insights to investigate properties of entangled operations.

\end{abstract}


\maketitle

Entanglement is the key concept that makes quantum physics distinct from classical physics and plays an essential role in quantum information processing. As an entangled state cannot factor into a product of individual states, entangled operations cannot be represented by a product of individual operations but a coherent superposition of operations~\cite{Zanardi01, Zanardi00, Dur01,Wang02, Nielsen03}. To date, superpositions of operators have been studied in both fundamental and practical aspects. Commutation relations for bosonic operators~\cite{Zavatta09} and Pauli operators~\cite{Yao10, kim10} have been directly demonstrated experimentally using superpositions of operators. Recently, it is reported that the superpositions of quantum operations can allow controlling the orders of quantum operations, which is not allowed in the traditional formalism of quantum physics~\cite{Oreshkov12, Chiribella13, Procopio15}. This, so-called {\it superposition of causal orders}, is not only fundamentally interesting but also proven that it has practical applications in quantum information processing in the context of communication complexity~\cite{Guerin16, Wei19}, witnessing causality~\cite{Rubino17, Goswami18}, quantum metrology~\cite{Zhao19}, and transmission of quantum information~\cite{Guo20}.

However, so far research on superposition of quantum operations has been limited to superposition of single qubit gates in most cases, meaning that entanglement of operators has not been studied much yet. Most of the implementations of superposition of causal orders are relying on the interferometric scheme, however, in this case extension to an entangled operation is not straightforward since nonlocality should be involved. Extension to implementing an entangled operation is particularly important since a two-qubit entangled operation such as controlled-NOT (CNOT) gate can constitute the universal sets of quantum gates with single qubit gates~\cite{nielsen02,Knill01,Barz15}. Implementation of entangled operations for two-qubit systems can provide further advantages for practical quantum information processing, such as entanglement filter~\cite{Hofmann02,Okamoto09,Zhou11}, entangling gate for universal quantum gate set~\cite{Lu19}, and entanglement generation for teleportation-based programmable quantum gate~\cite{Yao10, Nielsen97,Slodicka09,Watson18}. It is also intriguing to explore entangled operations in even larger Hilbert spaces for investigating potential applications in quantum information processing.

In this letter, we propose and demonstrate a scheme for realizing an arbitrary entangled operation based on a coherent superposition of local operations. We first begin by introducing the concept of operator entanglement using an operator-Schmidt decomposition. We then present some intriguing and important operations that can be realized based on this scheme in two-qubit systems. Then, we describe our experimental demonstration of several two-qubit entangled operations in photonic systems and discuss our experimental results. Finally, we extend our scheme to the case of generalized entangled operations with $N$ qubit systems and the Schmidt number $M$. We also present a practical scheme on implementing three-qubit entangled operations and introduce interesting three-qubit operations that can be implemented.


An operator $\mathcal{O}_{{\rm AB}}$ acting on systems A and B can be written as
\begin{equation}
\mathcal{O}_{{\rm AB}}=\sum_{i}c_i \mathcal{O}_{\rm A}^i \otimes \mathcal{O}_{\rm B}^i,
\label{eq:OSchmit}
\end{equation}
where $c_i \ge 0$ and  $\mathcal{O}_{\rm A}^i $ $(\mathcal{O}_{\rm B}^i )$ is orthonormal bases for a system A (B)~\cite{Nielsen03}. Similar to the Schmidt representation of quantum states, Equation~(\ref{eq:OSchmit}) is the operator-Schmidt decomposition and the number of nonzero Schmidt coefficients $c_i$ is defined as a Schmidt number~\cite{Nielsen03}. If the Schmidt number of an operator $\mathcal{O}_{{\rm AB}}$ is larger than one, $\mathcal{O}_{{\rm AB}}$ is an entangled operator meaning that it can generate entanglement from separable states. It is clear from Eq.~(\ref{eq:OSchmit}) that an entangled operation cannot be prepared by local operations and classical communications (LOCC) between two distant systems A and B.  In general, an operator having higher Schmidt number needs more resources to be constructed~\cite{Nielsen03}.

\begin{figure}[htbp]
\centering
\includegraphics[width=3.4in]{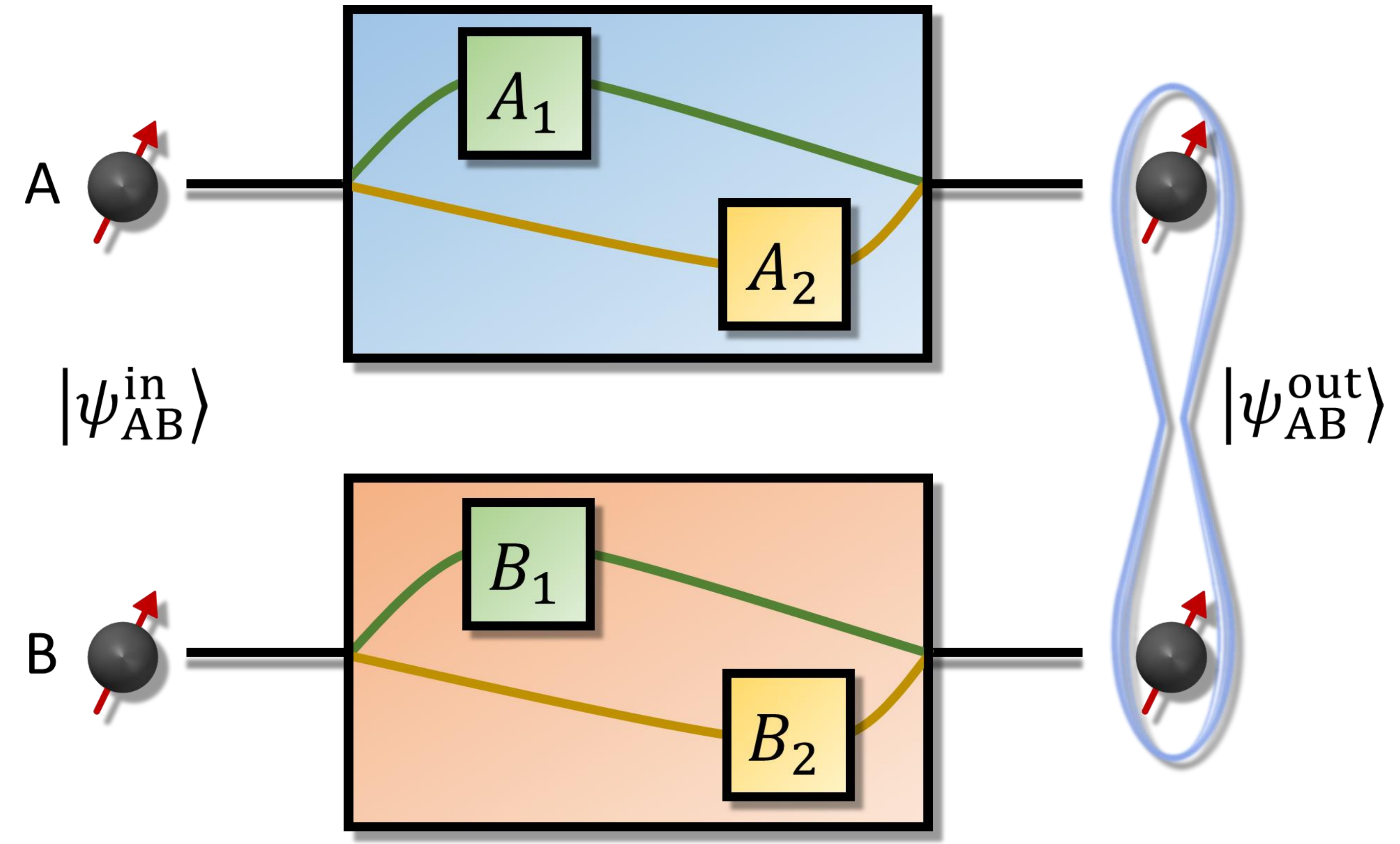} 
\caption{Concept of two-qubit entangled operation with the Schmidt number 2. Two-qubit input states $\left | \psi_{{\rm AB}}^{{\rm in}} \right \rangle$ undergoes either operation $A_1 \otimes B_1$ (green line) or $A_2 \otimes B_2$ (orange line). Here, these two processes are coherently superposed so that the output state is $\left | \psi_{{\rm AB}}^{{\rm out}} \right \rangle=\frac{1}{\sqrt{2}} \left (A_1 \otimes B_1 + e^{i \phi} A_2 \otimes B_2 \right )\left | \psi_{{\rm AB}}^{{\rm in}} \right \rangle$. For a separable input state, the output state can be entangled. Note that the entangled operation can be implemented only with local operations if we use additional degree of freedom (ancillary systems).} 
\label{fig:scheme}
\end{figure}

Let us consider a simple case with the Schmidt number 2 and $c_1=c_2$. Then, 
\begin{eqnarray}
\mathcal{O}_{{\rm AB}}&=&\frac{1}{\sqrt{2}} \left (\mathcal{O}_{\rm A}^1  \otimes \mathcal{O}_{\rm B}^1 + \mathcal{O}_{\rm A}^2  \otimes \mathcal{O}_{\rm B}^2 \right ) \nonumber \\ 
 &=& \frac{1}{\sqrt{2}} \left (A_1 \otimes B_1 + e^{i \phi} A_2 \otimes B_2 \right ),
\label{eq:EntOp}
\end{eqnarray}
where $\mathcal{O}_{\rm A}^1=A_1$, $\mathcal{O}_{\rm A}^2=A_2$, $\mathcal{O}_{\rm B}^1=B_1$, $\mathcal{O}_{\rm B}^2=e^{i \phi} B_2$ and $\phi$ is the relative phase between $A_1 \otimes B_1$ and $A_2 \otimes B_2$. Note that the second line of Eq.~(\ref{eq:EntOp}) is not the Schmidt decomposition due to the complex number of coefficients, but it can explicitly show the relative phase $\phi$ between two superposed operations. In many cases, local operations are fixed but their relative phase can be varied. The entangled operation of Equation (\ref{eq:EntOp}) has a simple form, however, there are intriguing operations. When $A_1=B_1=I$, $A_2=B_2=\sigma_x$, Equation (\ref{eq:EntOp}) becomes 
\begin{equation}
\frac{1}{\sqrt{2}}\left (I\otimes I+e^{i\phi}\sigma_x\otimes\sigma_x \right )=\frac{1}{\sqrt{2}}\left( \begin{array}{cccc}
1 & 0 & 0 & e^{i\phi}\\
0 & 1 & e^{i\phi} & 0\\
0 & e^{i\phi} & 1 & 0\\
e^{i\phi} & 0 & 0 & 1
\end{array}  \right),
\label{eq:EO1}
\end{equation}
where $I$ is an identity operation and $\sigma_x$, $\sigma_y$, and $\sigma_z$ are Pauli operators.  One can prepare maximally entangled states $\frac{1}{\sqrt{2}}(\left | 0 0 \right \rangle + e^{{i \phi}}\left | 11 \right \rangle)$ ($\frac{1}{\sqrt{2}}(\left | 0 1 \right \rangle + e^{{i \phi}}\left | 10 \right \rangle)$) from a separable input state $\left | 0 0 \right \rangle$ ($\left | 0 1 \right \rangle$). 

When $A_1=\left | 0 \right \rangle \left \langle 0 \right |$, $B_1=I$, $A_2=\left | 1 \right \rangle \left \langle 1 \right |$, and $B_2=U$ where $U$ is a single qubit unitary operation, one can implement a controlled-Unitary (CU) operation.  In addition, by setting $A_1=B_1=|0\rangle\langle 0|$, $A_2=B_2=|1\rangle\langle 1|$, one can realize the entanglement filter (EF) which is a special non-unitary operation, and transmits the input states only when the incoming qubit states is either $|0\rangle |0\rangle $ or $|1\rangle |1\rangle $~\cite{Hofmann02,Okamoto09,Zhou11}.

However, it is difficult to implement an entangled operation of Equation (\ref{eq:EntOp}). Using an interferometer, it is possible to make a superposition of two different operators such as $\mathcal{O}_{\rm A}= 1/\sqrt{2} \left (A_1 +e^{i \phi_A}  A_2 \right )$~\cite{kim10}. However, when the system A and B use interferometers separately, what they prepare is $1/2 (A_1 + e^{i \phi_A}A_2) \otimes (B_1 + e^{i \phi_B} B_2 )$, which has clearly the Schmidt number 1.  Note that if we apply $A_1$ and $B_1$ simultaneously to the system A and B with the half of the time and apply $A_2$ and $B_2$ with another half of the time, the prepared operation is an incoherent mixture of $A_1 \otimes B_1$ and $A_2 \otimes B_2$ rather than a coherent superposition of them.


Here, we propose a scheme for implementing an entangled operations using additional degree of freedom. The conceptual diagram of our scheme is shown in Fig.~\ref{fig:scheme}. For a two-qubit intput state $\left | \psi_{{\rm AB}}^{{\rm in}} \right \rangle$, qubit A undergoes an operation $A_1$ or $A_2$ while qubit B undergoes an operation $B_1$ or $B_2$. In order to prepare an entangled operation, one needs to rule out $A_1 \otimes B_2$ and $A_2 \otimes B_1$ cases and the other two cases $A_1 \otimes B_1$ and $A_2 \otimes B_2$ should be coherently superposed. It is difficult to achieve these requirements, however, it is possible by using energy-time correlation of spontaneous parametric down conversion (SPDC) process. In experiment, the input state is encoded in a polarization of single photons. When a pair of photons are generated via SPDC process, the generated photon pairs have strong energy-time correlation. By exploiting this strong time correlation between two photons, one can prepare a coherent superposition of local operations, e.g., $\frac{1}{\sqrt{2}} \left (A_1 \otimes B_1 + e^{i \phi} A_2 \otimes B_2 \right )$ acting on polarization states of two-photons. 

\begin{figure*}[htbp]
\centering
\includegraphics[scale=0.28]{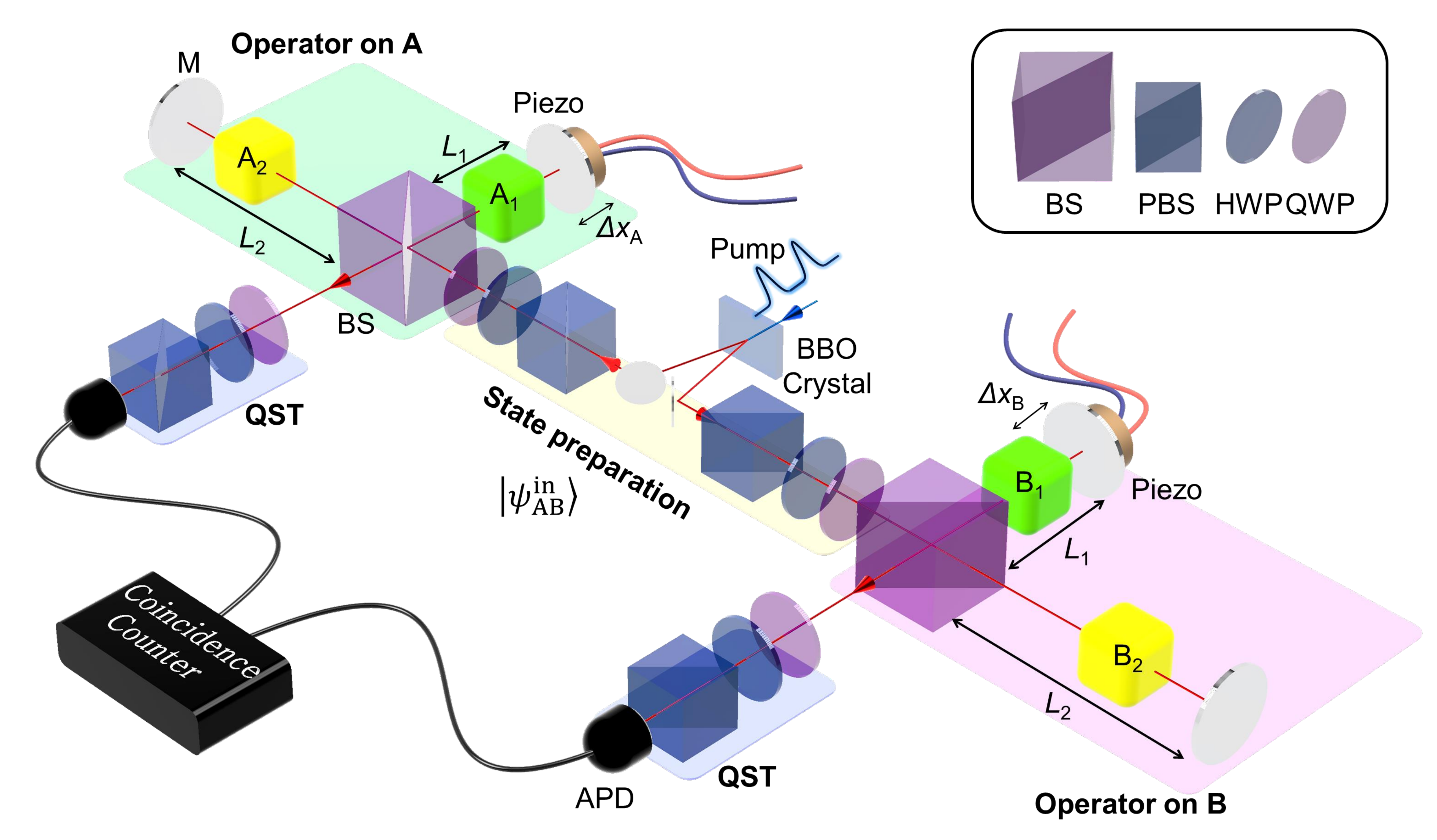}
\caption{Schematic of the experimental set-up. A down-converted photon pair generated from a pulsed laser is prepared in an arbitrary polarization two-photon state $|\psi_{{\rm AB}}^{{\rm in}}\rangle$ using a set of Pol., HWP, and QWP. Single qubit operations ($A_1$, $A_2$, $B_1$, and $B_2$) can be realized with linear optical elements such as Pol. and WPs, and the relative phase between two arms of an unbalanced Michelson interferometer (UMI) is controlled by adjusting $\Delta x_{\rm A}$ and $\Delta x_{\rm B}$, respectively.  Entangled operations are implemented with two UMIs. A set of Pol. and WPs in front of APD are used to perform QST measurements (BBO: Beta Barium borate crystal, QST: Quantum state tomography, M: Mirror, APD: Avalanche photo diode, QWP: Quarter waveplate, HWP: Half waveplate, Pol.: Polarizer, BS: Beam splitter, PBS: Polarizing beam splitter).}
\label{fig:setup}
\end{figure*}

The schematic of our experimental setup is shown in Fig.~\ref{fig:setup}, which is based on Franson interferometer~\cite{Franson91, Thew04}. We use a femtosecond pulsed laser operating at a center wavelength of 780 nm and a pulse period of 12.5 ns. The wavelength of the pump laser becomes 390 nm using a lithium triborate (LBO) crystal via second harmonic generation process. Then, a pair of photons is generated via type-II SPDC process by pumping a 1 mm-thick BBO crystal, and its polarization state is prepared in an arbitrary input state $| \psi_{{\rm AB}}^{{\rm in}}\rangle$ with a set of polarizer, half and quarter wave plates (WPs). Then, each photon is sent to an unbalanced Michelson interferometer (UMI) with different optical lengths of $L_1$ and $L_2$. The path length different $L=L_2-L_1$ is 1.875 m corresponding to an half of the pulse period of the pump laser to ensure that the down-converted photons generated from consecutive pump pulses are temporally overlapped. Here we use an interference filter with 2 nm full-width at half maximum bandwidth, then the coherence length of each down-converted photon is around 300 ${\rm \mu}$m. Hence, there is no first order interference at the output of UMI. In order to implement $\mathcal{O}_{{\rm AB}}$ in Equation (\ref{eq:EntOp}), single qubit operations $A_1$ and $A_2$ ($B_1$ and $B_2$) are located in $L_1$ and $L_2$ arms of the UMI for a photon A (photon B), respectively. Note that a single qubit operation can be realized by linear optical elements such as WPs, polarizers, and so on. The two-photon output states are analyzed by a set of WPs and polarizers using quantum state tomography (QST) and quantum process tomography (QPT).

There are four possible output cases for the two-photon input states: $A_1\otimes B_1 \left |\psi_{{\rm AB}}^{{\rm in}} \right \rangle$, $A_1\otimes B_2 \left |\psi_{{\rm AB}}^{{\rm in}} \right \rangle$, $A_2\otimes B_1 \left |\psi_{{\rm AB}}^{{\rm in}} \right \rangle$, and $A_2\otimes B_2 \left |\psi_{{\rm AB}}^{{\rm in}} \right \rangle$. However, if we set the coincidence window (we use 3 ns) smaller than the time difference between $L_1$ and $L_2$ (12.5 ns), we can post-select only $A_1\otimes B_1 \left |\psi_{{\rm AB}}^{{\rm in}} \right \rangle$ and $A_2\otimes B_2 \left |\psi_{{\rm AB}}^{{\rm in}} \right \rangle$ events. Note that this is possible due to strong time correlation between generated photon pairs. Therefore, the post-selected output state $\left | \psi_{{\rm AB}}^{{\rm out}} \right \rangle$ becomes $\frac{1}{\sqrt{2}} \left(A_1\otimes B_1+e^{i\phi} A_2\otimes B_2 \right)  \left |\psi_{{\rm AB}}^{{\rm in}} \right \rangle$ where $\phi$ is the sum of relative phase differences in each UMI $(\phi = \phi_A + \phi_B)$. In order to fix $\phi$, we attached piezoelectric actuators on both mirrors in the short arms of the UMI and drive them using proportional integral derivative (PID) controllers to lock the relative phases $\phi_{\rm A}$ and $\phi_{\rm B}$. Here, we can choose $\phi$ between 0 to $2 \pi$ using a set of QWP, HWP, and QWP on the locking laser (780 nm pump laser). See Appendix for detailed information on the controllable phase-locking technique we use.

\begin{figure*}[htbp]
\centering
\includegraphics[scale=0.37]{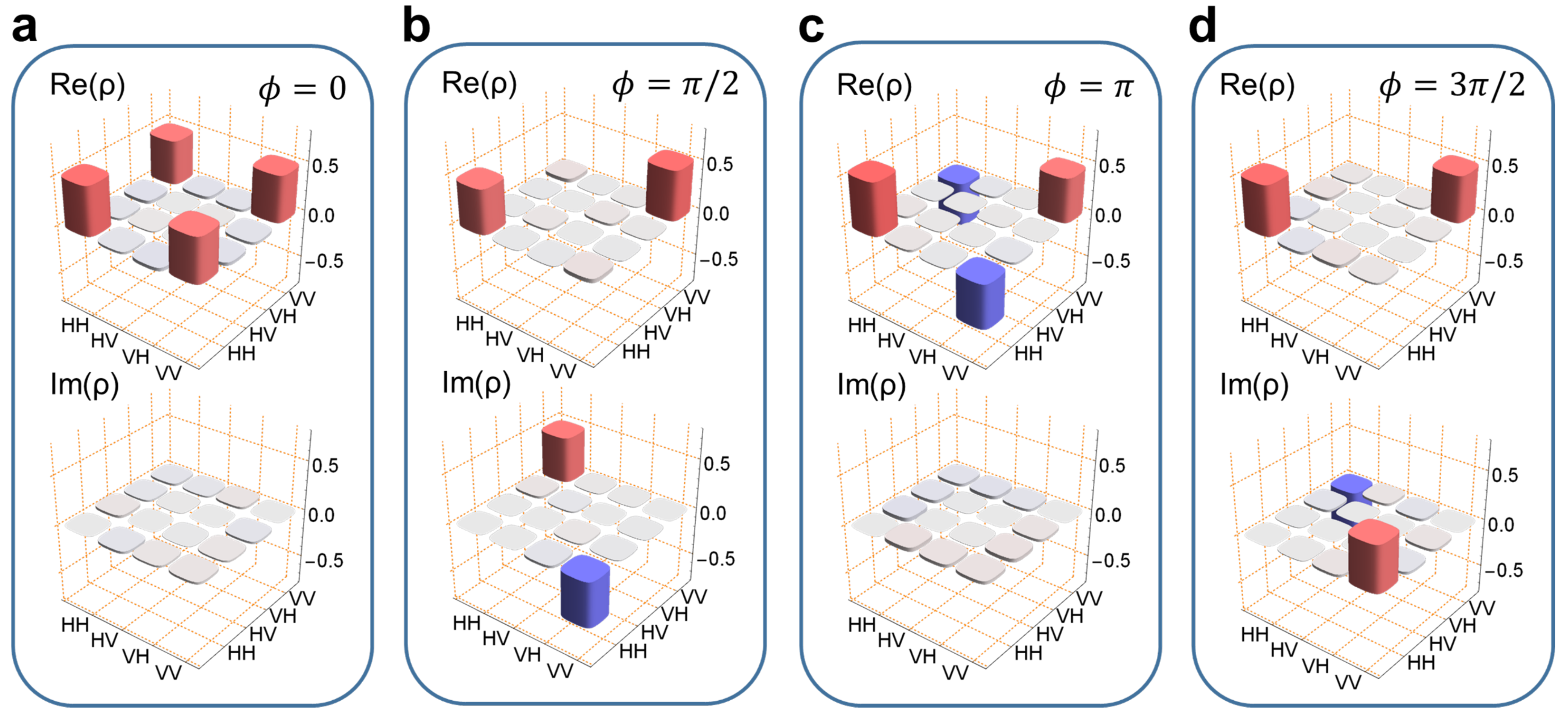} 
\caption{Reconstructed density matrices of the output state after the entangled operation $\mathcal{O}_{{\rm AB}}=\frac{1}{\sqrt{2}}\left (\sigma_z\otimes\sigma_z+e^{i\phi}\sigma_x\otimes\sigma_x \right)$. Input state is prepared to $|HH\rangle$, and $\sigma_x$ and $\sigma_z$ are implemented by using QWP with an angle of $\pi/4$ and $0$, respectively. The expected output state after the entangled operation is $\frac{1}{\sqrt{2}} \left(|HH\rangle+e^{i\phi}|VV\rangle \right)$. The reconstructed density matrices of the output states correspond to (a) $\phi$ = $0$, (b) $\pi/2$, (c) $\pi$, and (d) $3\pi/2$. Note that the upper (lower) rows corresponds to the real (imaginary) part of the density matrices. The average fidelity between the ideal output state and the experimentally obtained output state is $F_{{\rm ave}} = 0.940\pm 0.014$ and the average concurrence of the output states $C = 0.906\pm 0.025$. We performed Monte-Carlo simulations on reconstructing density matrices of each output state 100 times and the calculated experimental errors corresponds to one standard deviation.}
\label{fig:QST}
\end{figure*}

We demonstrated various entangled operations based on our experimental setup shown in Fig.~\ref{fig:setup}. The first entangled operation we consider is $\mathcal{O}_{{\rm AB}} (\phi)$ with  $A_1=B_1=\sigma_z$ and $A_2=B_2=\sigma_x$, i.e., $\mathcal{O}_{{\rm AB}}(\phi) = \frac{1}{\sqrt{2}} \left( \sigma_z \otimes \sigma_z + e^{i\phi}  \sigma_x \otimes \sigma_x \right)$. $\mathcal{O}_{{\rm AB}}(\phi)$ is able to generate entanglement since for a separable input state $\left |H H \right \rangle$, the output state is $\left | \psi _{{\rm AB}}^{{\rm out}} \right \rangle=\frac{1}{\sqrt{2}} \left (|HH\rangle+e^{i\phi}|VV\rangle \right )$, the maximally entangled states. We implemented $\mathcal{O}_{{\rm AB}}(\phi)$ with $\phi = 0, \pi/2, \pi$, and $3\pi/2$ and performed quantum state tomography (QST) measurement to each output states. The density matrices of the output states are reconstructed using maximum-likelihood method~\cite{James01,Fiurasek01} and the experimental results are shown in Fig.~\ref{fig:QST}. As shown in Fig.~\ref{fig:QST}, one can adjust the relative phase $\phi$ and the realized operation works well. Note that for the input state $\left |H V \right \rangle$, the output state becomes $\left | \psi _{{\rm AB}}^{{\rm out}} \right \rangle=\frac{1}{\sqrt{2}} \left(|HV\rangle-e^{i\phi}|VH\rangle \right )$. See Appendix for experimental results for the input state $\left |H V \right \rangle$. 

\begin{figure*}[htbp]
\centering
\includegraphics[scale=0.36]{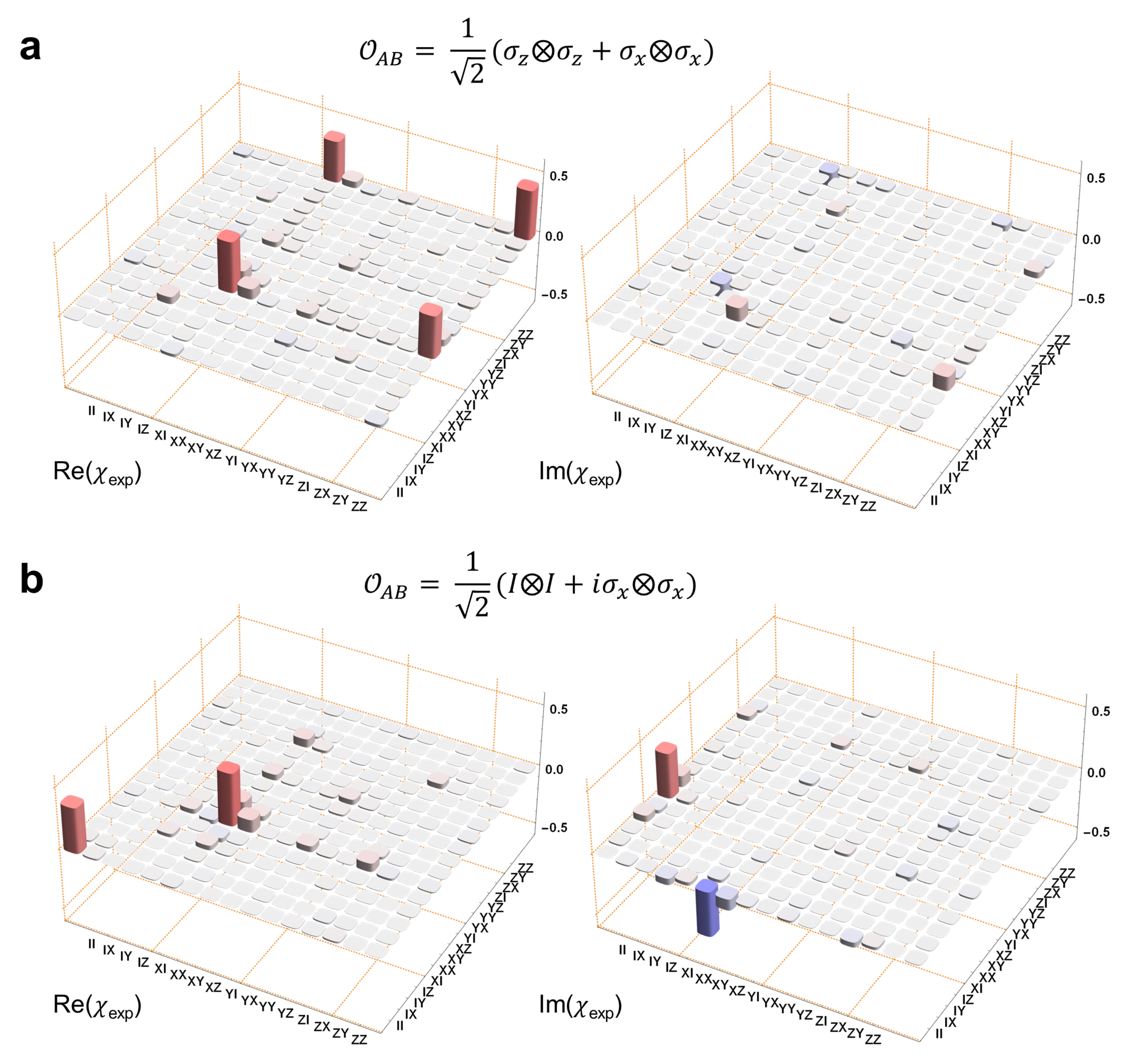} 
\caption{ Experimentally reconstructed process matrices $\chi$ of the entangled operations. (a). $\mathcal{O}_{{\rm AB}}=\frac{1}{\sqrt{2}} \left(\sigma_z\otimes\sigma_z+\sigma_x\otimes\sigma_x \right)$. (b) $\mathcal{O}_{{\rm AB}}=\frac{1}{\sqrt{2}} \left(I\otimes I + i \sigma_x\otimes\sigma_x \right)$. We reconstructed process matrices using maximum-likelihood method on quantum process tomography results. Left (right) columns corresponds to the real (imaginary) part of the process matrices. Process matrices are represented in the Pauli basis, for example, YZ corresponds to $\sigma_y$, $\sigma_z$ basis.}
\label{fig:QPT}
\end{figure*}

In order to show the quality of the implemented operation, we also performed quantum process tomography (QPT) measurement. Figure~\ref{fig:QPT} shows the experimentally reconstructed process matrices $\chi$ for entangled operations $\frac{1}{\sqrt{2}} \left (\sigma_z\otimes\sigma_z+\sigma_x\otimes\sigma_x \right)$ and $\frac{1}{\sqrt{2}} \left (I\otimes I + i\sigma_x\otimes\sigma_x \right )$. The corresponding process fidelities $F_{\chi}$ are $0.760 \pm 0.005$ and $0.762 \pm 0.006$, respectively. Here we use $F_{\chi} = \left[ {\rm Tr} (\sqrt{\sqrt{\chi_{{\rm exp}}} \chi_{{\rm ideal}}  \sqrt{\chi_{{\rm exp}}} }) \right ]^2$ where $ \chi_{{\rm ideal}}$ ($ \chi_{{\rm exp}}$) is the process matrix of the ideal (experimentally realized) operation. Note that $F_{\chi}$ results are obtained by performing 100 Monte-Carlo simulations on the QPT measurement results and the errors on the process fidelities corresponds to one standard deviation. We also provide QPT results of $\frac{1}{\sqrt{2}} \left (I\otimes I - \sigma_x\otimes\sigma_x \right)$ operation in Appendix. We attribute the non-unity process fidelity to the imperfections of our experimental setup. We provide performance test results of our phase stability setup in Supplementary Note. We emphasize that non-unity process fidelity is not due to the limitation of our proposed scheme but the imperfection of our experimental demonstration. 

Note that in general, a coherent superposition of unitary operators is not a unitary operator. The operator $\frac{1}{\sqrt{2}} \left (\sigma_z\otimes\sigma_z+\sigma_x\otimes\sigma_x \right)$ is composed of unitary operators but it is a non-unitary operator. On the other hand, $\frac{1}{\sqrt{2}} \left (I\otimes I + i\sigma_x\otimes\sigma_x \right)$ is a unitary operation. It is known that if $U_{{\rm AB}}$ is a two-qubit unitary operation with Schmidt number 2, $U_{{\rm AB}}$ has the form of  $\sqrt{1-p} I \otimes I + i \sqrt{p} \sigma_x \otimes \sigma_x$ up to local unitary equivalence where $ 0\le p \le 1$~\cite{Nielsen03}. Hence, one can realize any two-qubit unitary operations with Schmidt number 2 using our scheme. Moreover, it is interesting to consider the possibility of emulating Ising gate, which is a two-qubit gate implemented natively in trapped-ion quantum system, based on our scheme~\cite{Jones03, Debnath16}. For example, $\frac{1}{\sqrt{2}} \left (I\otimes I + i\sigma_x\otimes\sigma_x\right)$ is exactly the same with $3\pi/4$ Ising (XX) gate~\cite{Debnath16}. Note that, Ising (XX) gate and single-qubit rotation gates can constitute a universal set of quantum gates.

\begin{figure}[htbp]
\centering
\includegraphics[width=3.4in]{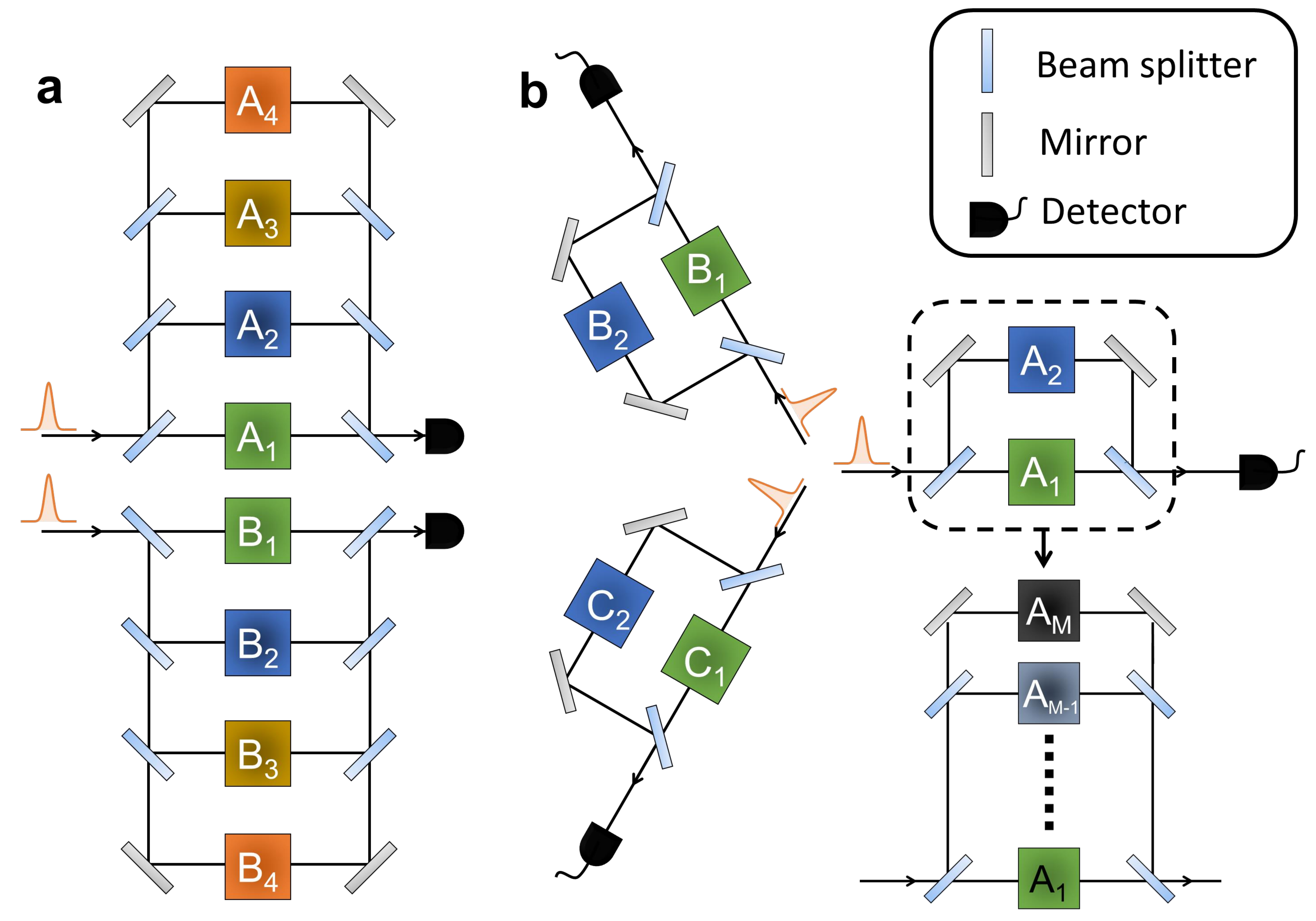} 
\caption{Generalization to the entangled operation with higher Schmidt number and multipartite systems. (a) For two-qubit input states, one can increase the number of optical paths of the UMZI. Then, the entangled operation with higher Schmit number can be realized. (b) By adding one more UMZI, one can implement tripartite entangled operations.}
\label{fig:general}
\end{figure}

We emphasize that entangled operations can be extended to the case of $M$ superposed operations $\mathcal{O}_N^M$ for $N$ qubit states by adding optical arms of the interferometers and adding the number of interferometers, respectively, then $\mathcal{O}_N^M$ is given by
\begin{equation}
\mathcal{O}_N^M=\sum\limits_{k=1}^{M} c_k \mathcal{O}^{(1)}_k \otimes \ldots  \otimes \mathcal{O}^{(N)}_k,
\end{equation}
where $\mathcal{O}^{(j)}_k$ is the $k$-th local operation acting on $j$-th qubit and $c_k$ is the Schmidt coefficient. Figure~\ref{fig:general}(a) shows an unbalanced Mach-Zehnder interferometer (UMZI) with four different arms.  For an interferometer A, the local operation $A_i$ with $i=0, 1, 2, 3$ is located in each arm of the UMZI where the optical path length difference is fixed to $L$ between two neighboring arms and it is the same for the interferometer B. Then, one can prepare an entangled operation with Schmit number 4 of 
\begin{equation}
\frac{1}{2}\left (A_1\otimes B_1+e^{i\phi_1}A_2\otimes B_2+e^{i\phi_2}A_3\otimes B_3+e^{i\phi_3}A_4\otimes B_4 \right),
\label{eq:fourqubit}
\end{equation}
by post-selecting the case where the arrival time difference of two-photons is zero and $\phi_i$ with $i=1,2,3$ is the relative phase differences. In this case, all the other terms are not detected~\cite{Franson91, Thew04}. It is well-known that any two-qubit operation can be decomposed into the form of Eq.~(\ref{eq:fourqubit}) up to local unitary equivalence~\cite{Kraus01}. Note that we can implement SWAP gate $\frac{1}{2}\left (I\otimes I+\sigma_x\otimes\sigma_x+\sigma_y\otimes\sigma_y+\sigma_z\otimes\sigma_z \right)$ based on the setup shown in Fig.~\ref{fig:general}(a). Considering the CNOT complexity of SWAP gate is three~\cite{Coffey08}, i.e., three consecutive CNOT gates are required for a single SWAP gate, it significantly reduces the implementation complexity of SWAP gate.

We can also consider increasing the number of qubits by adding more UMZIs as shown in Fig.~\ref{fig:general}(b). For example, three photon input states can be prepared by using cascaded SPDC process~\cite{Agne17}. In this case, we consider three qubit systems and the corresponding three qubit entangled operation $\mathcal{O}_3^M$ is
\begin{equation}
\mathcal{O}_3^M = \sum\limits_{k=1}^{M} c_k \mathcal{O}_k^{\rm A}\otimes \mathcal{O}_k^{\rm B} \otimes \mathcal{O}_k^{\rm C},
\end{equation}
where $\mathcal{O}_k^{\rm C}$ are single qubit operations acting on qubit C. For example, one can obtain three qubit Greenberger-Horne-Zeilinger (GHZ) state~\cite{Greenberger89} $\left | GHZ \right \rangle = \frac{1}{\sqrt{2}}\left(\left | H H H  \right \rangle+ \left | V V V  \right \rangle \right)$ from an input state $\left | H H H  \right \rangle$ by implementing $\frac{1}{\sqrt{2}} \left (I \otimes I \otimes I + \sigma_x \otimes \sigma_x \otimes \sigma_x \right )$. Likewise, if one implement $\frac{1}{\sqrt{3}} \left (I \otimes I \otimes \sigma_x + I \otimes \sigma_x \otimes I +  \sigma_x \otimes I \otimes I \right )$, then one can prepare three qubit W state~\cite{Dur00} $\left | W \right \rangle = \frac{1}{\sqrt{3}}(\left | H H V  \right \rangle+ \left | H V H  \right \rangle+ \left | V H H  \right \rangle )$ from an input state  $\left | H H H  \right \rangle$.  Moreover, one can implement the controlled-controlled-unitary (CCU) gate $\mathcal{O}_{{\rm CCU}}=|H\rangle \langle H | \otimes |H\rangle \langle H | \otimes I+|H\rangle \langle H | \otimes |V\rangle \langle V | \otimes I+|V\rangle \langle V | \otimes |H\rangle \langle H | \otimes I+|V\rangle \langle V | \otimes |V\rangle \langle V | \otimes U$. If $U=\sigma_x$, then this operation becomes the Toffoli (controlled-controlled NOT) gate~\cite{Lanyon09}. So far, we only consider superposition of local operations with equal amplitudes but the amplitude of each operation can be varied by adjusting transmittance and reflectance of beam splitters.


In summary, we have proposed a scheme to implement arbitrary entangled operations based on a coherent superposition of local operations. We also report an experimental demonstration of various two-qubit entangled operations in photonic systems. We believe that our scheme has great importance both in fundamental and practical aspects. Entanglement of operators is closely related to the evolution of quantum systems under nonlocal Hamiltonians and the entangling power of quantum operators~\cite{Dur01, Zanardi00, Alba19}. In addition, it is known that superposition of quantum gates cannot be represented by the conventional quantum circuit models~\cite{Arujo14} and the extended quantum circuit model allowing such superposition of operators can reduce the computational complexity of some problems and simplify realization of various quantum information processing~\cite{Arujo14B, Chiribella13}. Due to the simplicity of our scheme and the possibility to extend to multi-qubit systems, it is interesting to apply our scheme to solve practical problems using quantum algorithms such as quantum chemistry problem~\cite{Cao19}. Moreover, although we mainly discuss unitary operations but the scheme can also be applicable to non-trace-preserving operations. Hence, it is intriguing to consider exploring PT symmetry broken operations~\cite{Klauck19} and adding a control to weak measurements or partial collapse measurement~\cite{Lim14}.   







\newpage

\section{Appendix}
\subsection{Experimental details}
\begin{figure*}[htbp]
\centering
\includegraphics[width=5.8in]{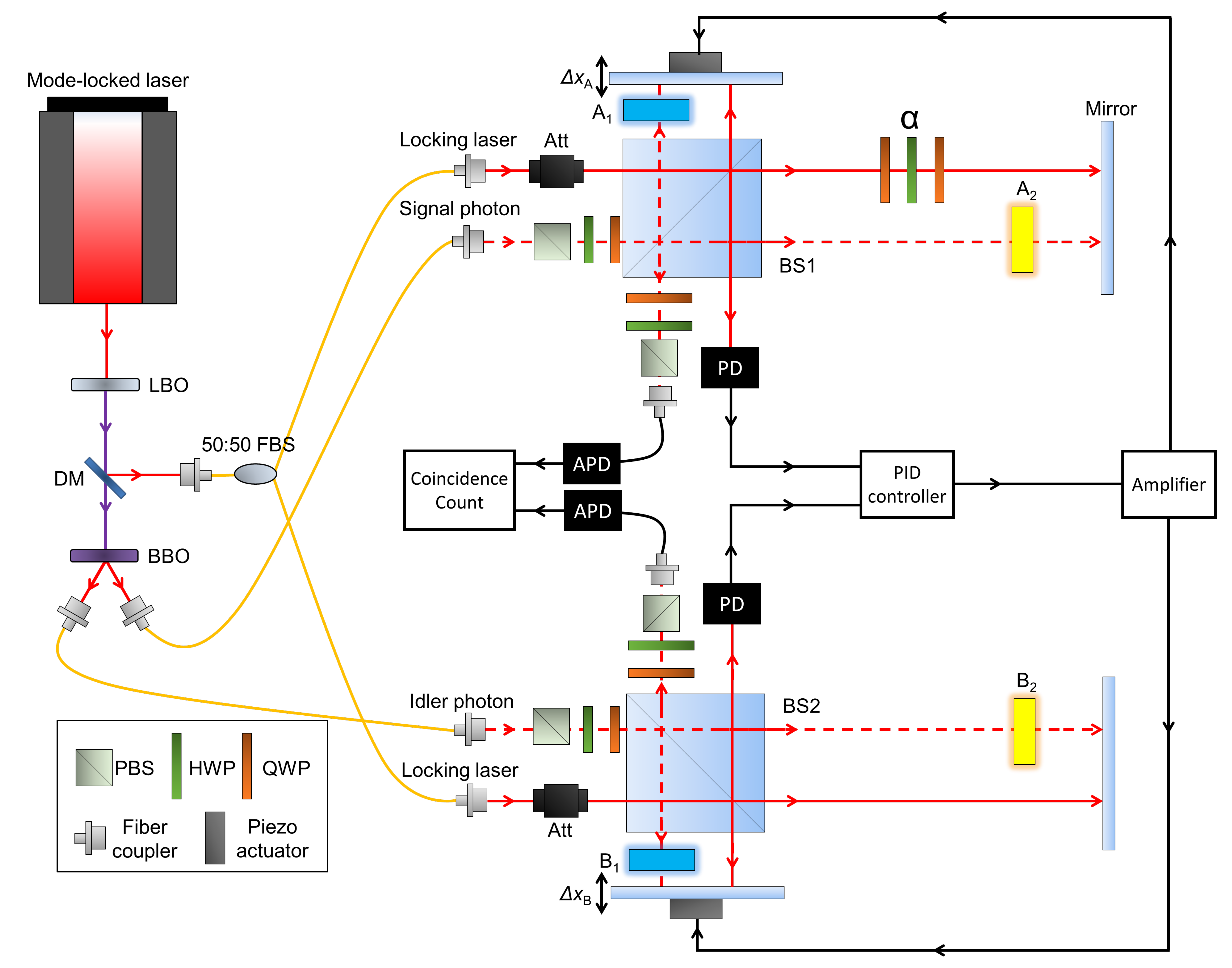} 
\caption{Experimental setup for implementing two-qubit entangled operation $\frac{1}{\sqrt{2}}\left( A_1 \otimes B_1 + e^{i\phi}A_2 \otimes B_2 \right )$. LBO: Lithium triborate, DM: dichroic mirror, FBS: fiber beam splitter, BBO: beta-barium borate, BS: beam splitter, PBS: polarizing beam splitter, HWP: half wave plate, QWP: quarter wave plate, PID: proportional-integral-derivative, PD: photo detector, APD: avalanche photo diode, Att: attenuator.}
\label{Sfig:setup}
\end{figure*}

390 nm pulse train is prepared by 1 mm thick Lithium triborate (LBO) crystal via second-harmonic generation (SHG) process using a 780 nm mode-locked pump laser (80 MHz repetition rate, 140 fs pulse duration, $M^2< 1.1$, Chameleon, Coherent), and generates photon pairs via type-II spontaneous parametric down-conversion (SPDC) process at a 1 mm thick beta-barium borate (BBO) crystal. The 780 nm pump laser is separated by a dichroic mirror (DM) from 390 nm laser, and used as a locking laser to lock the relative phase between two arms of an unbalanced Michelson interferometer (UMI). The locking laser enters a one-inch cube beam splitter (BS) spatially separated from the single photon's path in order not to be detected at APD. In order to actively lock the relative phase between two arms of UMI from mechanical and thermal vibrations of optical components, we implement a proportional-integral-derivative (PID) control system as shown in Fig.~\ref{Sfig:setup}. 

\subsection{Locking the relative phase between two arms of an unbalanced Michelson interferometer}
\begin{figure*}[htbp]
\centering
\includegraphics[width=6in]{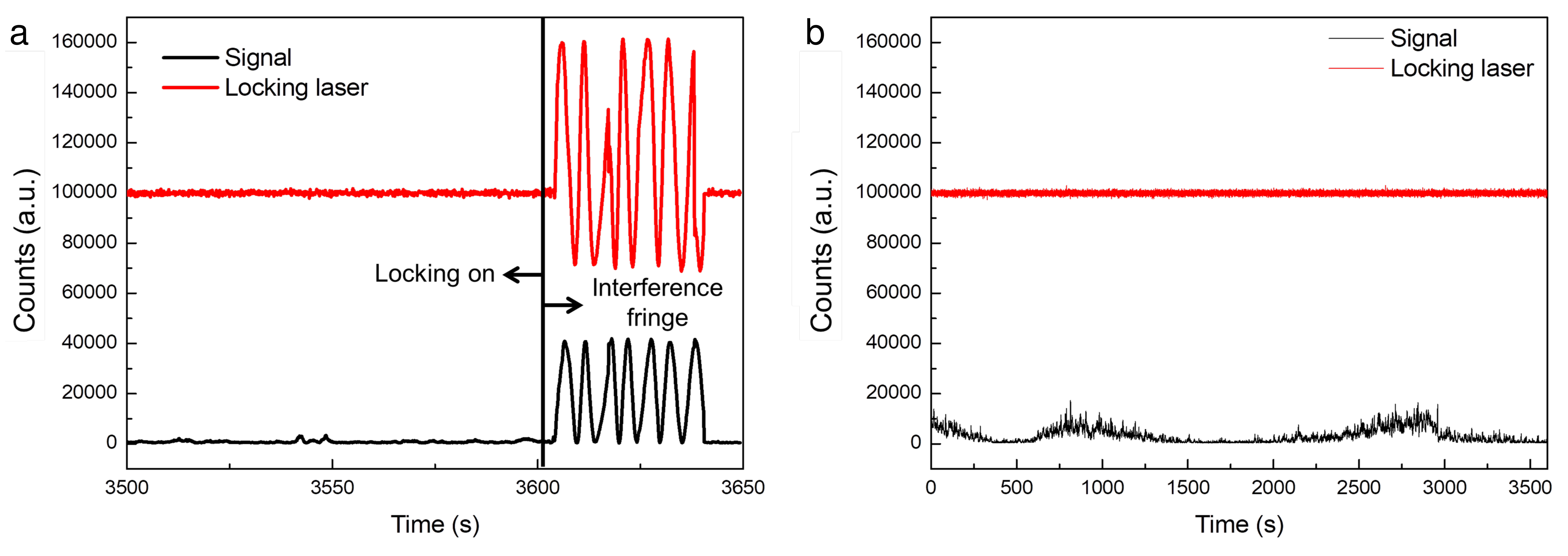} 
\caption{ Results on active phase locking of an Michelson interferometer. The relative phase of an balanced MI is locked by fixing the intensity of locking laser to around the half of the maximum intensity, and we find that single photon signal is also locked simultaneously. Red (black) line corresponds to the intensity of the locking laser (single photon signals). (a) Correlation between the locking laser intensity and the single photon signal. They are locked at the same time during phase locking is on. We can observe that interference fringes of the locking laser and the signal simultaneously by applying a periodic voltage to the piezo actuator. (b) We monitor the intensity change of the locking laser and the signal during an hour while the locking system is on.}
\label{Sfig:lock}
\end{figure*}
We test the performance of our phase locking system using a single balanced Michelson interferometer (MI), meaning that the path length difference between two arms is negligible. Single photon and locking laser are injected to a single balance MI, and the intensity of the locking laser is controlled by driving the voltage to a piezo actuator attached on the mirror of one arm. Then, we measure the single photon count of the signal (down-converted photons) using an APD depending on the intensity of the locking laser. As shown in Fig.~\ref{Sfig:lock}(a), we can observe the interference fringes of both signal (black line) and locking laser (red line) at the same time by scanning the mirror on one arm using a piezo actuator (See Fig.~\ref{Sfig:lock}(a) after 3600 s). Then, we fixed the intensity of the locking laser around the half of the maximum intensity of the interference fringe by using a PID controller and a voltage amplifier, which enables locking the relative phase between two arms of MI. As shown in Fig.~\ref{Sfig:lock}(a), signal is locked to have the minimum count by adjusting the phase using $\alpha$ when intensity of the locking laser is half of the maximum intensity of interference fringe. Figure~\ref{Sfig:lock}(b) shows the stability of phase locking in the interferometer during an hour, we find that the relative phase on signal is locked. We have considered that the imperfections of our experimental results (purity, concurrence, and fidelity) are mainly due to the phase fluctuation. We implement an entangled operation $\frac{1}{\sqrt{2}}\left( A_1 \otimes B_1 + e^{i\phi}A_2 \otimes B_2 \right )$ using a single UMI instead of two UMIs and the results shows the better performance. We will compare this results at the Section III. 

In order to control the relative phase $\phi$ of $\frac{1}{\sqrt{2}}\left( A_1 \otimes B_1 + e^{i\phi}A_2 \otimes B_2 \right )$ from 0 to $2\pi$ in two UMIs, we use a set of QWP($\pi/4$)-HWP($\alpha$)-QWP($\pi/4$) for locking laser on the long arm of the UMI on qubit A, see Fig.~\ref{Sfig:setup}. This waveplate combination provides the phase retardation $\phi = 4\alpha+\pi$~\cite{Brendel95} depending on the HWP's angle $\alpha$ while the polarization is unchanged. Note that, this set of WPs gives a phase shift only for locking laser (not signal). If we give phase difference between two arms of UMI by changing the HWP's angle $\alpha$, the intensity will be changed according to phase difference from $\alpha$. However, the optical path length difference is shifted depending on $\alpha$ while keeping the intensity of locking laser at half maximum of the interference fringe using PID control, and it enables a relative phase to be adjustable to any value from 0 to $2\pi$. 

\begin{figure*}[htbp]
\centering
\includegraphics[width=6.0in]{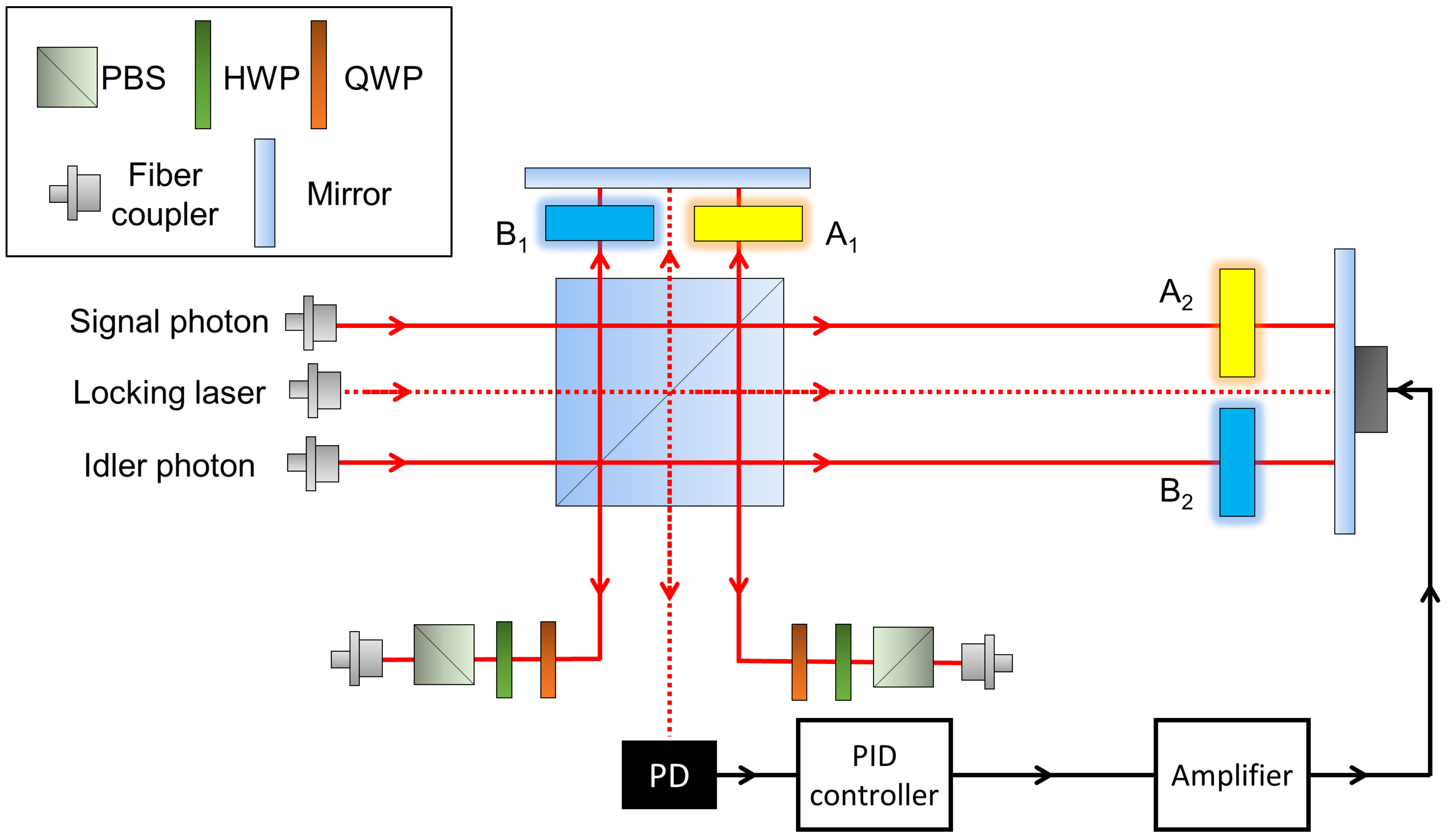} 
\caption{Experimental setup for implementing $\frac{1}{\sqrt{2}}\left( A_1\otimes B_1+ e^{i\phi}A_2\otimes B_2 \right )$ entangled operation using a single UMI. This corresponds to the simplified version of the setup shown in  Fig.~\ref{Sfig:setup}. Note that the number of UMI is reduced from two to one, meaning that phase locking is required for only one interferometer.}
\label{Sfig:setup2}
\end{figure*}

\subsection{Experimental setup with a single UMI instead of two}

\begin{figure*}[htbp]
\centering
\includegraphics[width=6.6in]{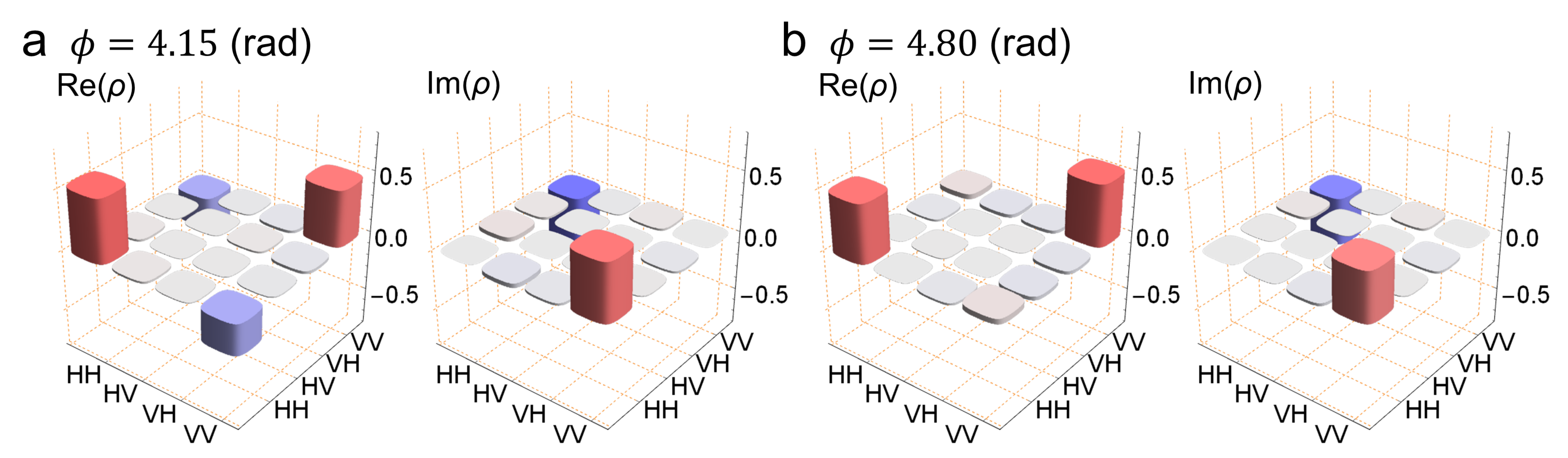} 
\caption{ Experimental QST results with single UMI setup shown in Fig.~\ref{Sfig:setup2}. The realized entangled operation is $\frac{1}{\sqrt{2}}\left( \sigma_z\otimes\sigma_z+e^{i\phi}\sigma_x\otimes\sigma_x \right )$ and the input state is $\left | H H \right \rangle$. (a) $\phi = 4.15$ and (b) $\phi = 4.80$.}
\label{Sfig:QST}
\end{figure*}

In order to confirm that imperfections of our experimental results are mainly from our phase locking system on the UMIs, we realize the same experiment using a single UMI~\cite{Sanaka01,Rossi08} rather than two UMIs as shown in  Fig.~\ref{Sfig:setup2}. Here we send the signal and idler photons into a single BS with a locking laser. Then we can lock the phase of a single UMI instead of two using a single locking laser beam. In this setup, we carry out quantum state tomography of the output state for the entangled operation $\frac{1}{\sqrt{2}}\left( \sigma_z\otimes\sigma_z+e^{i\phi}\sigma_x\otimes\sigma_x \right )$ and the results for the input state $\left | HH \right \rangle$ is shown in Fig.~\ref{Sfig:QST}. We compare the experimental results for two cases: a single UMI and two UMI setups in Table~\ref{table:UMI}. By decreasing the number of phase locking from two to one, the experimental results are improved. Purity, fidelity, and concurrence of the output state increase about 0.038, 0.027, and 0.039, respectively. Nevertheless, we have to uses two UMI setup of Fig.~\ref{Sfig:setup} due to the limited spaces for mounting various waveplates required to perform quantum state tomography (QST) and quantum process tomography (QPT). However the comparison summarized in Table~\ref{table:UMI} suggests that reducing number of interferometer can increase the quality of the realized operation. Thus, we believe that our phase locking system have room for improvement. 

\begin{table}[htbp]
\centering
\begin{tabular}{c c c}
\hline
\hline
 & 1 UMI & 2 UMIs \\ \hline
Purity & $0.942\pm0.018$ & $0.904\pm0.025$ \\ 
Fidelity & $0.967\pm0.010$ & $0.940\pm0.015$ \\
Concurrence & $0.945\pm0.017$ & $0.906\pm0.025$ \\   \hline \hline

\end{tabular}
\caption{Comparison of experimental results obtained with the setups consisting of either single or double UMIs.}
\label{table:UMI}
\end{table}

\subsection{Additional data of QST and QPT}
\begin{figure*}[htbp]
\centering
\includegraphics[width=6.6in]{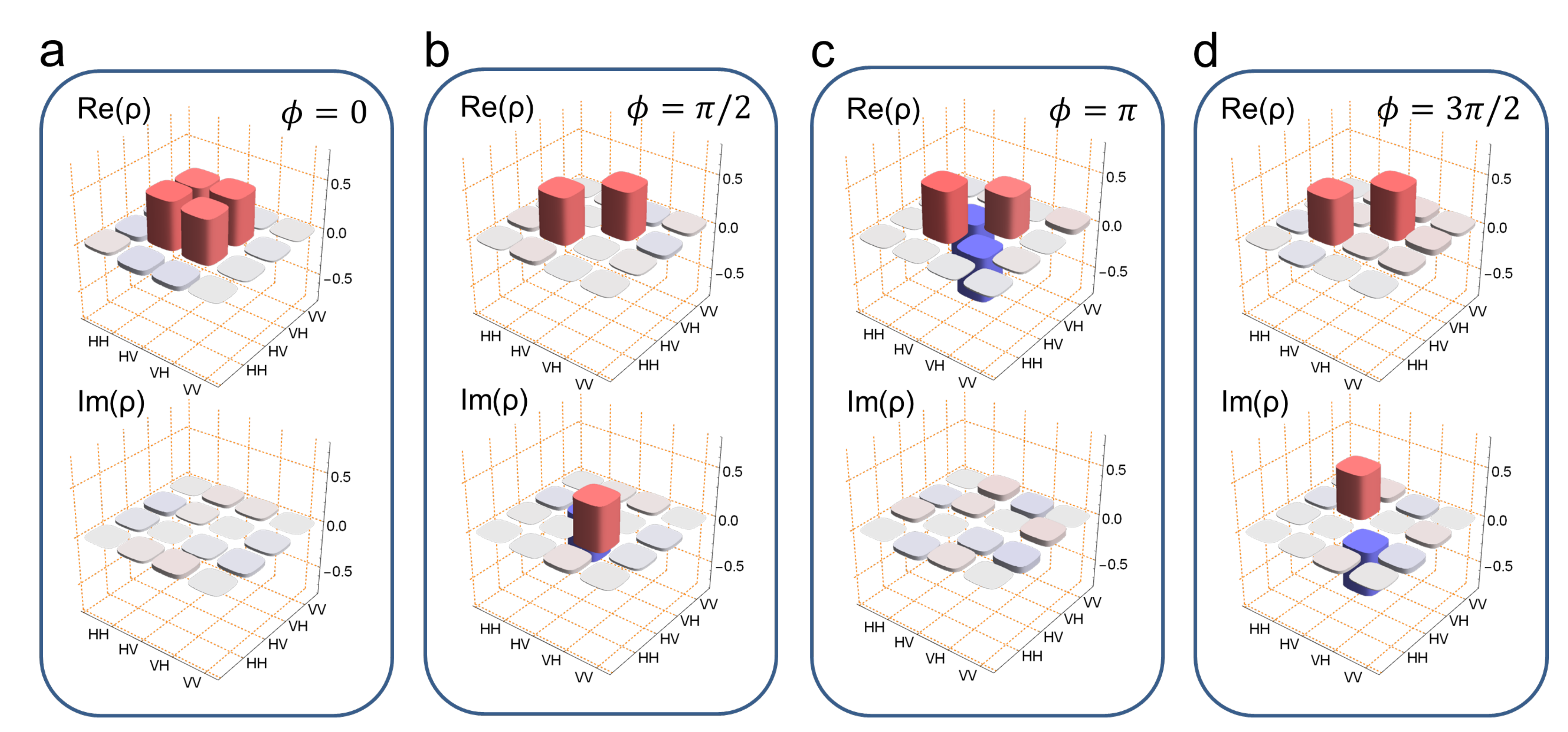} 
\caption{Reconstructed density matrices of the output state after the entangled operation $\mathcal{O}_{{\rm AB}}=1/\sqrt{2}(\sigma_z\otimes\sigma_z+e^{i\phi}\sigma_x\otimes\sigma_x)$. Input state is prepared to $|HV\rangle$and the expected output state after the entangled operation is $1/\sqrt{2}(|HV\rangle+e^{i\phi}|VH\rangle)$. The reconstructed density matrices of the output states correspond to (a) $\phi$ = $0$, (b) $\pi/2$, (c) $\pi$, and (d) $3\pi/2$. Note that the upper (lower) rows corresponds to the real (imaginary) part of the density matrices. The average fidelity between the ideal output state and the experimentally obtained output state is $F_{{\rm ave}} = 0.933\pm 0.010$ and the average concurrence of the output states $C = 0.904\pm 0.018$. We performed Monte-Carlo simulations on reconstructing density matrices of each output state 100 times and calculated experimental errors. Here the errors on each quantity corresponds to one standard deviation.}
\label{Sfig:QST2}
\end{figure*}

In this section, we provide additional experimental data that are not included in the main text. At first, Fig.~\ref{Sfig:QST2} shows the reconstructed output states after applying $1/\sqrt{2}(\sigma_z\otimes\sigma_z+e^{i\phi}\sigma_x\otimes\sigma_x)$ entangled operation and the expected output state is $1/\sqrt{2}(|HV\rangle+e^{i\phi}|VH\rangle)$ for the input state $\left| HV \right \rangle$. See Figure 3 of the main text for comparison. 

Furthermore, we perform quantum process tomography measurement on $\frac{1}{\sqrt{2}}\left( I\otimes I-\sigma_x\otimes\sigma_x \right )$ entangled operation and the experimentally reconstructed process matrix $\chi_{{\rm exp}}$ is shown in Supplementary Figure~\ref{Sfig:QPT}. The process fidelity of $\chi_{{\rm exp}}$ with respect to the ideal operation is $0.830\pm0.008$, which is obtained by performing Monte-Carlo simulations 100 times on the experimental data and the error corresponds to one standard deviation.

\begin{figure*}[htbp]
\centering
\includegraphics[width=6.6in]{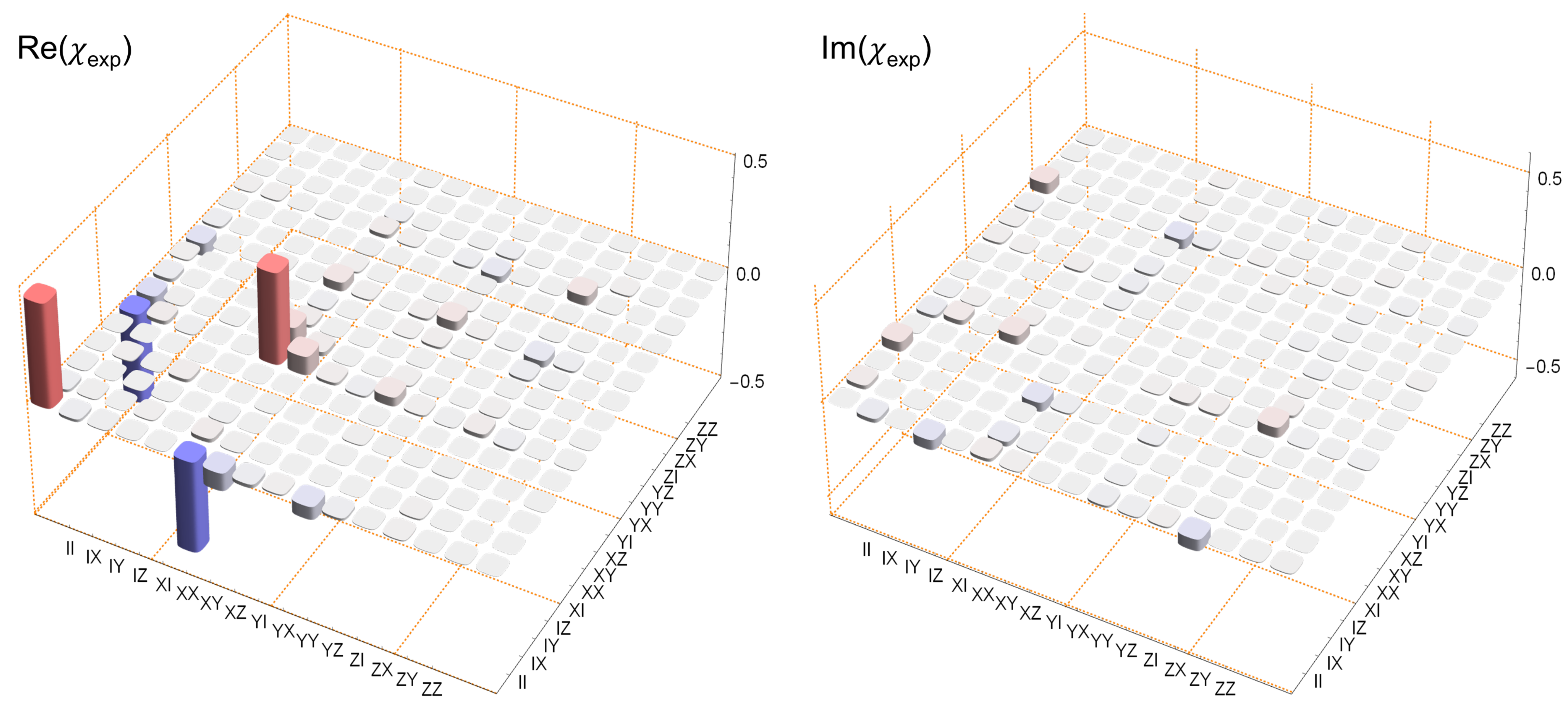} 
\caption{ Experimentally reconstructed process matrices $\chi$ of the entangled operation $\frac{1}{\sqrt{2}}\left( I\otimes I-\sigma_x\otimes\sigma_x \right )$. Left (right) columns corresponds to the real (imaginary) part of the process matrices. Process matrices are represented in the Pauli basis.}
\label{Sfig:QPT}
\end{figure*}

\end{document}